\documentclass[12pt]{iopart}

\usepackage{graphicx}
\usepackage{subfig}
\usepackage{amssymb}
\usepackage{amsthm} 
\usepackage{mathptmx}
\usepackage{nicefrac}
\usepackage{xcolor}
\usepackage[switch]{lineno} 

\begin{document}

\title[]{High precision X-ray spectroscopy of kaonic neon\footnote{We dedicate this work to the memory of Prof. C. Guaraldo and Prof. J. Zmeskal, whose contributions were essential to the success of the kaonic neon measurements. This work would not have been possible without them.}}

\author{ 
F Sgaramella$^{1*}$, D Sirghi$^{2,1,3,**}$, K Toho$^{4}$, F Clozza$^{1}$, L Abbene$^{5}$, C Amsler$^{6}$, F Artibani$^{1,7}$, M Bazzi$^1$, G Borghi$^{8,9}$, D Bosnar$^{10}$, M Bragadireanu$^{3}$, A Buttacavoli$^{5}$, M Cargnelli$^{6}$, M Carminati$^{8,9}$, A Clozza$^1$, R Del Grande$^{11,1}$, L De Paolis$^1$, K Dulski$^{1,12,13}$, L Fabbietti$^{11}$, C Fiorini$^{8,9}$, I Fri\v{s}\v{c}i\'c$^{10}$, C Guaraldo$^1,^\dagger$, M Iliescu$^1$, M Iwasaki$^{14}$, A Khreptak$^{12,13}$, S Manti$^1$, J Marton$^{6}$, P Moskal$^{12,13}$, F Napolitano$^1$, S Nied\'{z}wiecki$^{12,13}$, H Ohnishi$^{4}$, K Piscicchia$^{2,1}$, F Principato$^{5}$, A Scordo$^{1}$, M Silarski$^{12}$, F Sirghi$^{1,3}$, M Skurzok$^{12,13}$, A Spallone$^1$, L G Toscano$^{8,9}$, M T\"uchler$^{6}$, O Vazquez Doce$^1$, E Widmann$^6$, J Zmeskal$^{6,^\dagger}$ and C Curceanu$^1$}

\address{$^1$ Laboratori Nazionali di Frascati INFN, Frascati, Italy}
\address{$^2$ Centro Ricerche Enrico Fermi – Museo Storico della Fisica e Centro Studi e Ricerche “Enrico Fermi”, Roma, Italy}
\address{$^3$ Horia Hulubei National Institute of Physics and Nuclear Engineering (IFIN-HH) Măgurele, Romania}
\address{$^4$ Research Center for Accelerator and Radioisotope Science (RARiS), Tohoku University, Sendai, Japan}
\address{$^5$ Department of Physics and Chemistry (DiFC)—Emilio Segrè, University of Palermo, Palermo, Italy}
\address{$^6$ Stefan-Meyer-Institut f\"ur Subatomare Physik, Vienna, Austria}
\address{$^7$ Università degli studi di Roma Tre, Dipartimento di Fisica, Roma, Italy}
\address{$^8$ Politecnico di Milano, Dipartimento di Elettronica, Informazione e Bioingegneria, Milano, Italy}
\address{$^9$ INFN Sezione di Milano, Milano, Italy}
\address{$^{10}$ Department of Physics, Faculty of Science, University of Zagreb, Zagreb, Croatia}
\address{$^{11}$ Physik Department E62, Technische Universität München, James-Franck-Straße 1, 85748 Garching, Germany}
\address{$^{12}$ Faculty of Physics, Astronomy, and Applied Computer Science, Jagiellonian University, Krakow, Poland}
\address{$^{13}$ Center for Theranostics, Jagiellonian University, Krakow, Poland}
\address{$^{14}$ RIKEN, Tokyo, Japan}
\address{$^\dagger$ deceased}

\ead{$^*$ francesco.sgaramella@lnf.infn.it (Corresponding Author)}
\ead{$^{**}$ Diana.Laura.Sirghi@lnf.infn.it (Corresponding Author)}

\vspace{10pt}

\begin{abstract}
The high-precision kaonic neon X-ray transitions measurement performed by the SIDDHARTA-2 collaboration at the DA$\Phi$NE collider is reported. Both the X-ray energies and yields for high-n transitions were measured, demonstrating the feasibility of sub-eV X-ray spectroscopy for kaonic atoms using low-Z gaseous targets. The measurement provides valuable insights into the de-excitation processes in kaonic atoms, providing new input data for the refinement of the corresponding theoretical models, and a framework for testing Quantum Electrodynamics in strange exotic atoms. 

\end{abstract}

\vspace{2pc}

\section{Introduction}
An exotic atom is an atomic system in which a negatively charged particle, other than an electron, is bound in an atomic orbit by its electromagnetic interaction with the nucleus. Predicted in 1940s by Tomonaga and Araki \cite{PhysRev.58.90.2}, the exotic atoms represent unique tools to probe the fundamental interactions at the low-energy frontier.\\
Depending on the nature of the particle bound to the nucleus, these exotic systems are ideal for a wide variety of investigations. Muonic atoms have been used for precise test of the QED \cite{Borie:1982ax} and electro-weak interaction \cite{Bernabeu:1974jt,Feinberg:1974nx}. On the other hand, since hadrons are much heavier than electrons and the Bohr radius depends on the inverse of the reduced mass of the atomic system, hadronic atoms feel the nuclear force for the low energy levels, and are used for the study of hadron-nucleus strong interaction. In this context, kaonic atoms are of particular interest because the kaon is the lightest hadron containing a strange quark. Up to now, the X-ray spectroscopy of kaonic atoms was exploited to investigate the low-energy strong interaction providing experimental data for a better understanding of the nature of the kaon-nucleon(s) interaction and of the related theoretical models \cite{SIDDHARTA:2011dsy,J-PARCE62:2022qnt}. On the other side, these systems represent an ideal test bench for QED. Muonic and antiprotonic atoms have recently proven to be of significant interest for the study of bound-state QED (BSQED) \cite{Paul:2020cnx}, since, for high-n transitions, the strong interaction contribution is negligible, and they can be considered as pure QED transitions. To date, BSQED has been investigated through few-electron highly charged ions (HCI) \cite{Indelicato:2019nij}; however, the uncertainties due to finite nuclear size effects and nuclear polarization or deformation limit the precision achievable with measurements of high-Z ions, where BSQED effects are most pronounced. In this context, transitions between circular Rydberg states in exotic atoms represent a promising alternative. Due to their compactness, even high-n energy transitions exhibit enhanced BSQED effects compared to their HCI counterparts \cite{Paul:2020cnx}.\\
To date, no theoretical calculations addressing BSQED effects in kaonic atoms have been reported. However, since the main contribution to BSQED is the vacuum polarization (VP), which scales linearly with the reduced mass of the system \cite{Adkins:2024wws}, it is possible to extrapolate the magnitude of BSQED effects for kaonic atoms and in particular for kaonic neon. Based on theoretical calculations performed for muonic and antiprotonic neon \cite{Paul:2020cnx}, where first-order QED contributions are approximately 0.4 eV and 100 eV, respectively, and second-order QED contributions range from meV to eV, we expect BSQED corrections for kaonic neon to be on the order of tens of eV for first-order QED effects and approximately 0.5 eV for second-order QED effects. The advantage of using kaonic atoms, such as kaonic neon, lies in their high reduced mass, which enhances the sensitivity of BSQED corrections compared to muonic atoms. While antiprotonic atoms exhibit larger BSQED effects compared to kaonic atoms, the fine structure effects in antiprotonic atoms require high energy resolution detectors to distinguish the level splittings caused by spin-orbit interactions. In contrast, kaonic atoms do not have fine structure contributions, making the experimental measurement and analysis simpler and more straightforward for studying BSQED effects.
Therefore, precision X-ray spectroscopy of kaonic atoms is complementary to ongoing muonic \cite{PhysRevLett.130.173001} and antiprotonic experiments, providing experimental data to refine and test BSQED predictions.\\
By pioneering development of a new generation of Silicon Drift Detectors (SDDs) for X-ray spectroscopy \cite{Miliucci:2021wbj}, with excellent energy and time resolutions and capable to work in the high radiation environment of particle accelerators, the SIDDHARTA-2 collaboration is performing high precision measurements of kaonic atoms transitions at the DA$\Phi$NE collider. In this work, we report the first measurement of kaonic neon X-rays, with several high-n transitions measured with sub-eV statistical error precision, providing results on both transition energies and yields, contributing to a deeper understanding of kaonic atoms cascades, paving the way for a new era of measurements aimed at precise QED tests.

\subsection{The SIDDHARTA-2 kaonic neon measurement}
\begin{figure}[htbp]
        \centering
        \includegraphics[width=0.6\textwidth]{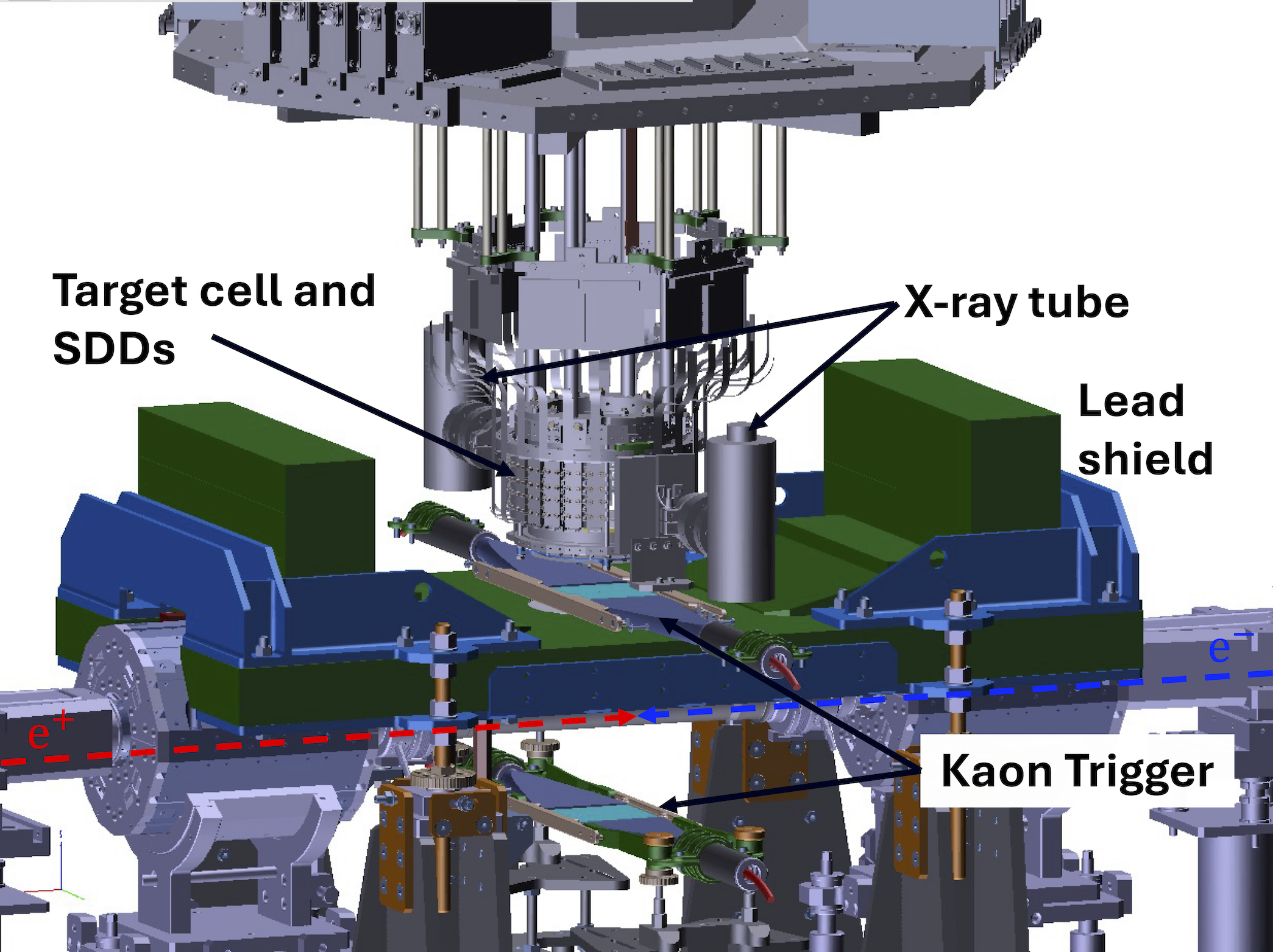}
        \caption{Schematic layout of the SIDDHARTA-2 experimental apparatus installed at the DA$\Phi$NE interaction point. The main elements, as the kaon trigger, the target cell surrounded by SDDs and the X-ray tubes for the detectors calibration, are highlighted.}
        \label{fig:setup}
\end{figure}

The SIDDHARTA-2 experiment, installed at the DA$\Phi$NE collider of the National Laboratories of Frascati (INFN-LNF) in Italy, is a cutting-edge experiment developed to perform high precision X-ray spectroscopy of kaonic atoms. The DA$\Phi$NE \cite{Milardi:2018sih,Milardi:2021khj,Milardi:2024efr} electron-positron collider delivers low momenta ($\sim$ 127 MeV/c) K$^+$K$^-$ pairs through the decay of the $\phi$ meson, being the ideal facility for the generation and study of kaonic atoms. The core of the experimental apparatus (Figure \ref{fig:setup}) consists of a cryogenic target cell made of a high-purity aluminum frame and 150 $\mu$m Kapton walls, ensuring high transparency for X-ray above 5 keV. The target is surrounded by 384 Silicon Drift Detectors (SDDs), with a total active area of 246 cm$^2$, developed specifically to perform X-ray spectroscopy. The very good energy resolution of these detectors ($\sim$150 eV FWHM at 6.4 keV \cite{Miliucci:2021wbj}), together with their excellent time resolution of 500 ns \cite{Miliucci:2022lvn} which allows to use them in coincidence with a kaon trigger for the background rejection, makes them ideal for spectroscopic measurements of kaonic atoms. The kaon trigger consists of two plastic scintillators, one installed above and the other below the beam pipe, to detect the back-to-back emitted K$^+$K$^-$ pairs. Coincident signals from both scintillators provide a clear identification of the K$^+$K$^-$ pairs, while rejecting events not synchronous with kaon production. More details on the setup are reported in \cite{Sirghi:2023wok}.\\
The energy calibration of the SDDs is crucial to guarantee the accuracy of the kaonic atoms measurements. It is performed using a system composed of two X-ray tubes and a multi-element target made of high-purity titanium, iron, and copper strips. The X-ray tubes induce the fluorescence emission in the target elements, and their characteristic lines are used to calibrate the SDDs. This procedure ensures an energy calibration accuracy within a few eV \cite{Sgaramella:2022rbl}, matching the requirements for precision X-ray spectroscopy.\\
The kaonic neon measurement was performed in Spring 2023, when the target cell was filled with neon gas and cooled down to 28 K to maintain a density of 0.3\% of the liquid neon density, corresponding to 3.6 g/l. The data were collected over approximately two weeks for a total integrated luminosity of 125 pb$^{-1}$.

\section{Data selection}
\begin{figure}[htbp]
        \centering
        \includegraphics[width=0.55\textwidth]{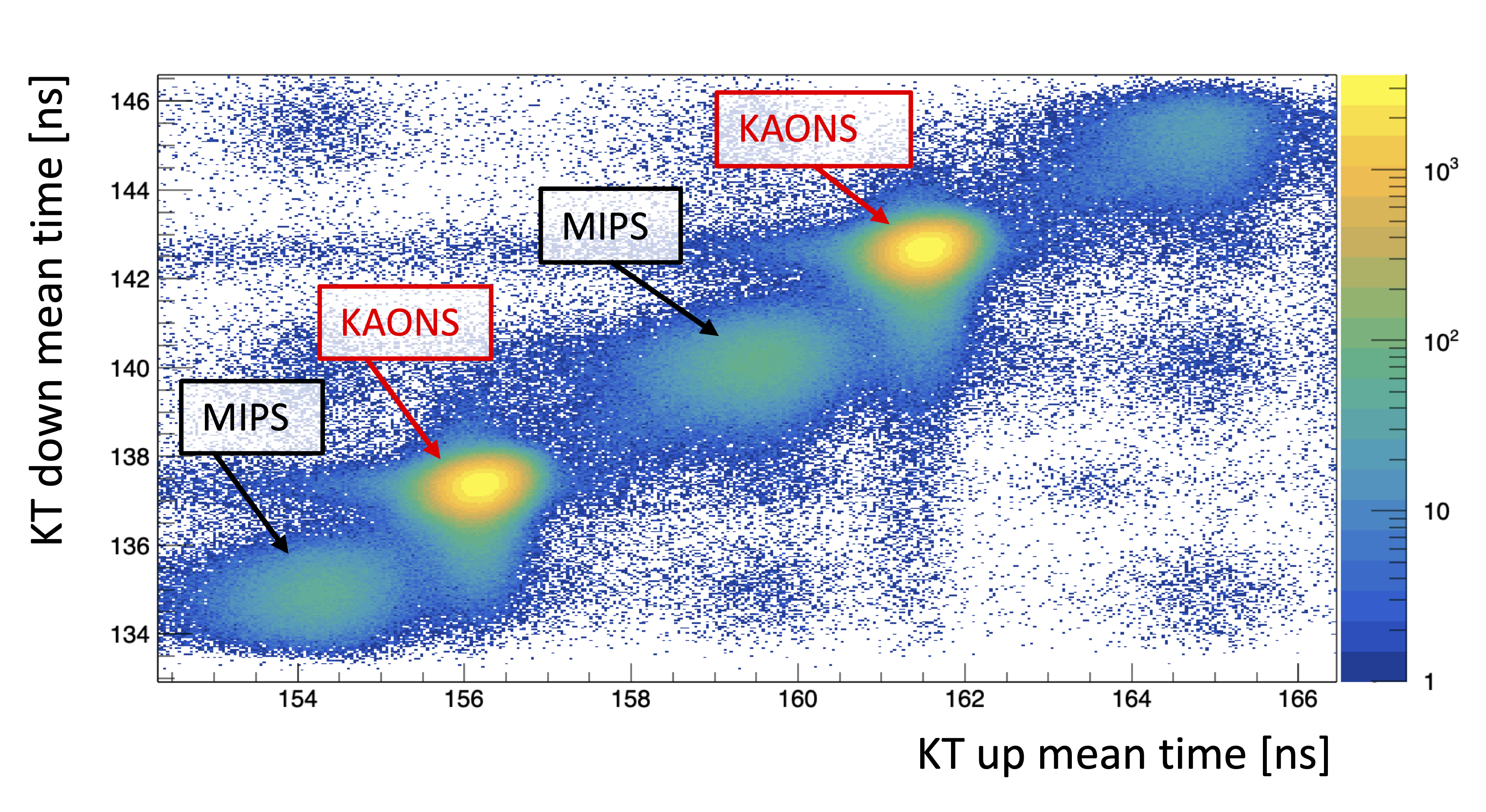}
        \includegraphics[width=0.4\textwidth]{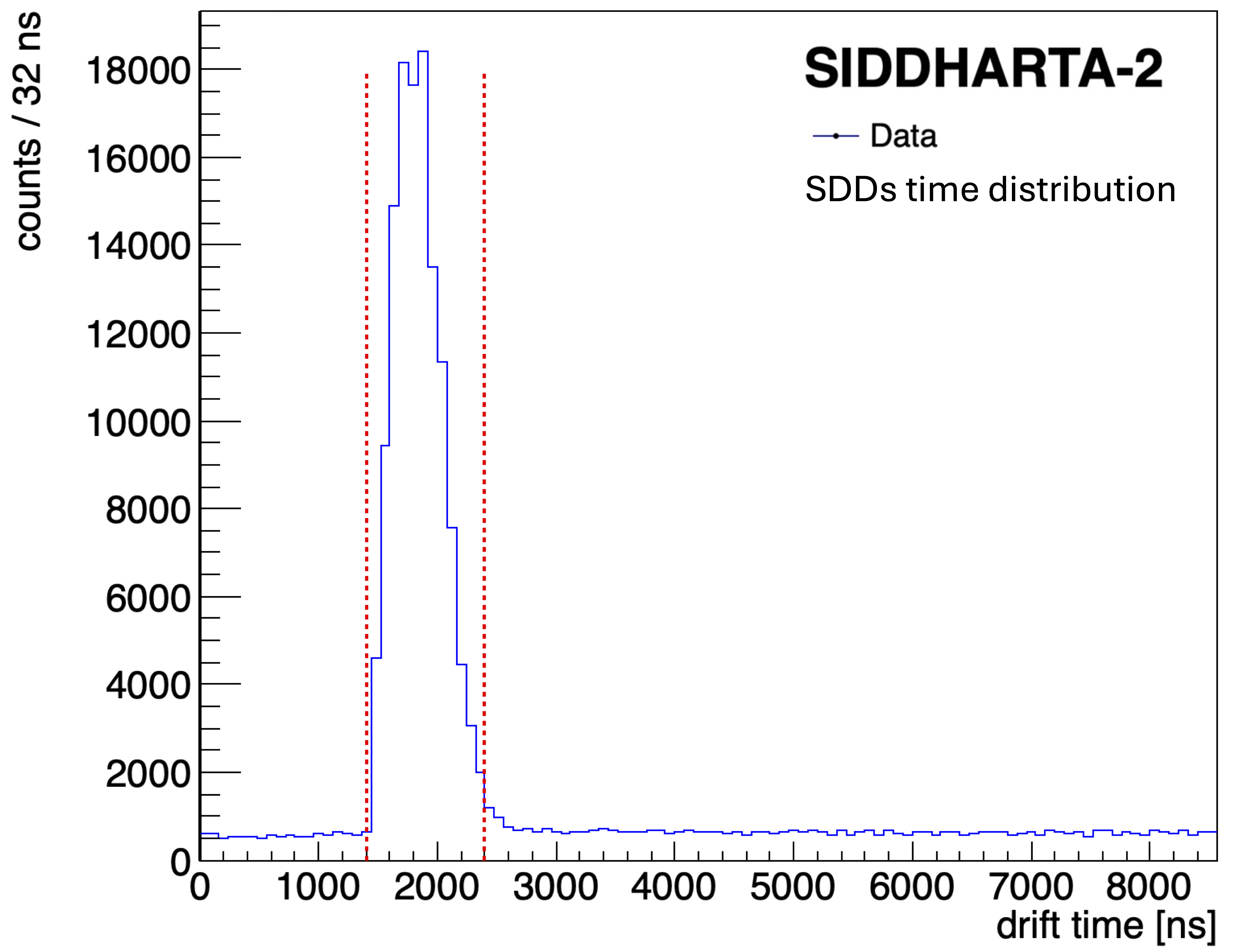}
        \caption{Left: Plot of the time difference between the kaon trigger top (KT up) and bottom (KT down) scintillators and the DA$\Phi$NE radiofrequency. The coincidence events related to kaons (high intensity) are clearly distinguishable from MIPs (low intensity). Right: Distribution of the time difference between the kaon trigger signals and X-ray hits on the SDDs. The dashed lines represent the 1.0 $\mu$s acceptance window.}
        \label{fig:KT}
    \end{figure}

An efficient selection of events is essential to disentangle the X-rays emitted by the kaonic neon de-excitation process from background-related events. The main source of background is represented by the electromagnetic showers induced by lost electrons and positrons, due to the Touschek effect and beam-gas interactions \cite{Boscolo:2011zz}. These events, which are asynchronous with the back-to-back K$^+$K$^-$ production, are rejected by the kaon trigger, which reduces the background by a factor of approximately 10$^4$ \cite{Sgaramella:2023orc}. However, minimum ionizing particles (MIPs) generated from beam-beam and beam-gas interactions can produce accidental trigger signals when passing simultaneously through the scintillators of the kaon trigger. To differentiate between K$^+$K$^-$ pairs and MIP-induced triggers, the time of flight technique is then employed. Since the kaons momentum is lower than that of MIPs, by measuring the time difference between the DA$\Phi$NE radiofrequency, which serves as a time reference, and the trigger signal, K$^+$K$^-$ pair events are disentangled from MIPs. Figure \ref{fig:KT}-left shows the correlation of the mean time distribution measured by the two scintillators of the kaon trigger during the kaonic neon run, demonstrating the effectiveness of the selection cut in distinguishing K$^+$K$^-$ pairs from MIPs. \\
The SDD time resolution plays a key role to enhance the background reduction. The time difference between the kaon trigger signals and the hits on the SDDs is shown in Figure \ref{fig:KT}-right. The peak corresponds to events in coincidence with the kaon trigger, while the flat distribution is due to uncorrelated events. By selecting the events falling within a 1.0 $\mu$s time window, determined by the SDDs' temporal resolution, the background is reduced by an additional factor about two \cite{Sgaramella:2023orc}.\\

\begin{figure}[htbp]
        \centering
        \includegraphics[width=1\textwidth]{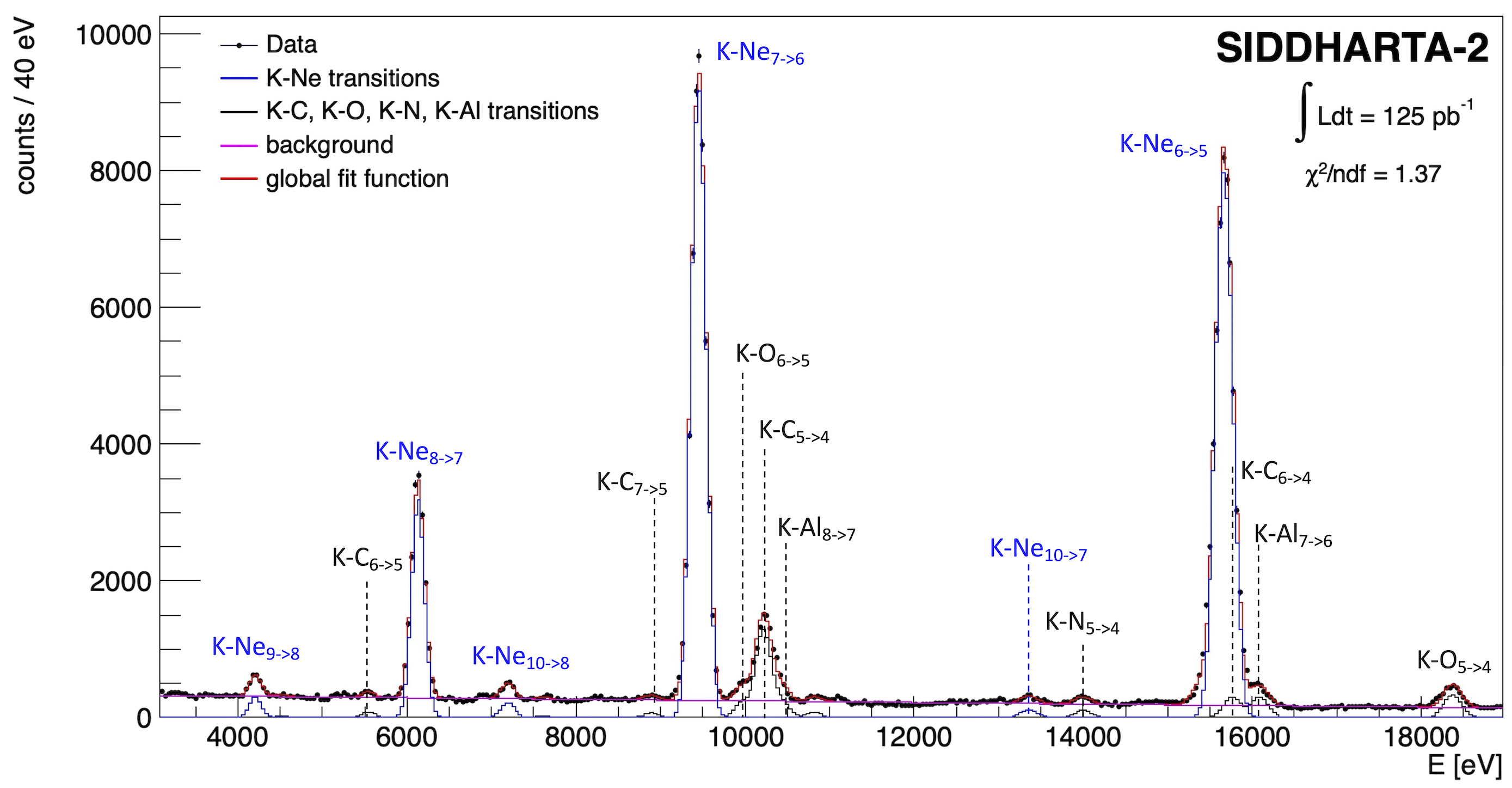}
        \caption{Kaonic Neon energy spectrum and relative fit after the events selection. The energy transitions are identified by the initial (n$_i$) and final (n$_f$) principal quantum numbers of the atomic levels. The several contributions of the fit function (red line) are highlighted: the kaonic neon (K-Ne) transitions in blue, the kaonic carbon (K-C), nitrogen (K-N), oxygen (K-O) and aluminium (K-Al) in black and the background in pink.}
        \label{fig:KNe_spectrum}
\end{figure}

\noindent
Figure \ref{fig:KNe_spectrum} shows the X-ray spectrum of the kaonic neon data obtained from the implementation of the event selection. Clear signals from kaonic atoms are observed, with highlighted peaks corresponding to X-ray emissions originating from kaonic atoms formed within the neon gas volume. The others lines are due to kaons stopped in the kapton (C$_{22}$H$_{10}$O$_5$N$_2$) entrance window and the aluminium frame of the target cell. To identify the kaonic neon peaks, the transition energies were calculated using the MCDFGME code \cite{Santos:2004bw}.

\section{Results and discussion}
\subsection{Kaonic neon transitions}
The energy associated with each kaonic atom transition, reported in Figure \ref{fig:KNe_spectrum}, was determined through spectral fitting technique. The energy response of the SDDs for each transition line is described by the convolution of a Gaussian with an exponential tail function to account for incomplete charge collection and electron-hole recombination effects \cite{Campbell:1990ye,CAMPBELL1997297,Gysel:2003}. A first-degree polynomial plus an exponential function were used to model the continuous background. The fitting was conducted over an energy range from 3 keV to 19 keV to include all observed kaonic neon lines. Six kaonic neon transitions were measured and their energy values are reported in Table \ref{tab:KNe}. Three of them, specifically the $8 \to 7$, $7 \to 6$, and $6 \to 5$ transitions, were determined with a statistical uncertainty below 1 eV. The systematic uncertainty is mainly limited by the calibration accuracy which could be improved in future experiments \cite{Curceanu:2023yuy}. 
These measurements provide new data for the kaonic atoms database and set a precedent for future high-precision kaonic atomic experiments, demonstrating the feasibility of sub-eV precision measurements using low-Z gaseous targets. This offers an advantage in terms of reduced electron screening effects and electron recapture, making high-n transitions in kaonic neon ideal for providing experimental input to investigate the BSQED.\\

    \begin{table}[htbp]
        \centering
        \caption{Kaonic neon energy transitions and absolute yields at the density of 3.60 $\pm$ 0.18 g/l. The first error is statistical, the second systematic.}
        \begin{tabular}{ccc}
            \hline
            \textbf{Transition} &   \textbf{Energy [eV]}  &   \textbf{Yield}  \\
            \hline
            K-Ne ($10 \to 8$) & $ 7191.21 \pm 4.91$ $\pm 2.00$ & $ \mathrm{ 0.010 \pm 0.001 \, \pm 0.001 \,} $ \\
            K-Ne ($10 \to 7$) & $13352.20 \pm 10.07$ $\pm 3.00$ & $ \mathrm{ 0.004 \pm 0.002 \, \pm 0.001 \,} $ \\ 
            K-Ne ($9 \to 8$) & $4206.35 \pm 3.75$ $\pm 2.20$ & $ \mathrm{ 0.137 \pm 0.012 \, \pm 0.010 \,} $ \\
            K-Ne ($8 \to 7$) & $6130.86 \pm 0.71$ $\pm 1.50$ & $ \mathrm{ 0.228 \pm 0.004 \, \pm 0.011 \,} $ \\
            K-Ne ($7 \to 6$) & $9450.08 \pm 0.41$ $\pm 1.50$ & $ \mathrm{ 0.277 \pm 0.002 \, \pm 0.014 \,} $ \\ 
            K-Ne ($6 \to 5$) & $15673.30\pm 0.52$ $\pm 9.00$ & $ \mathrm{ 0.308 \pm 0.003 \, \pm 0.015 \,} $ \\ 
            \hline
        \end{tabular}
        \label{tab:KNe}
    \end{table}

\subsection{Kaonic neon yields}
The X-ray yield for a given transition represents the probability that the kaonic atom will de-excite by emitting an X-ray corresponding to that specific transition. Measuring the yields of various transitions is crucial to understand the processes involved in the atomic cascade, since during the de-excitation, X-ray emission competes with other processes, such as Auger emission, Coulomb de-excitation, and, finally, nuclear absorption of the kaon. The development of theoretical models which fully describe the de-excitation process has been prevented by the lack of experimental data needed to constrain their free parameters. \\
Additionally, in high-precision QED tests with kaonic atoms, electrons play a crucial role due to their screening effect, which alters the energies of transitions. During the kaon cascade, some electrons are ejected via the Auger effect. This process not only depletes the atom of electrons but also reduces the transition yields, as the emission of an Auger electron suppresses the corresponding radiative transition. By comparing the measured yields with theoretical models that account for different numbers of remaining electrons, the number of bound electrons can be determined.\\
The absolute yield (Y) for a transition is defined by the ratio between the experimental detection efficiency ($\epsilon^{EXP}$) and the Monte Carlo efficiency ($\epsilon^{MC}$), and is given by: 
\begin{equation}
\centering
Y = \frac{\epsilon^{EXP}}{\epsilon^{MC}} = \frac{N_{X-ray}^{exp}/N_{KT}^{exp}}{N_{X-ray}^{MC}/N_{KT}^{MC}}
\label{eq:yield}
\end{equation}
The Monte Carlo simulation is based on GEANT4 \cite{Allison:2016lfl,Allison:2006ve,GEANT4:2002zbu}, reproducing the data taking conditions including the trigger efficiency, the interaction of the kaons with the neon gas and the efficiency of the SDDs \cite{Sirghi:2023scw}. The experimental efficiency $\epsilon^{EXP}$ is obtained from the number of detected X-ray normalized to the number of kaon triggers ($N_{KT}$), while $\epsilon^{MC}$ is given by the number of simulated X-ray, assuming a yield of 100\%, normalized to the number of simulated triggered kaon.\\

\noindent
In the initial phase of the kaonic neon run, the configurations of the target cell and kaon trigger have been optimized to maximize the signal-to-background ratio. To extract the yields of the transitions, we used the data corresponding to an integrated luminosity of 36 pb$^{-1}$ collected under stable conditions, allowing for a comparison with the Monte Carlo simulation. The results are shown in Table \ref{tab:KNe} with statistical uncertainties given by the fit. The systematic error is mainly due to the experimental uncertainty in the gas density and the thickness of the materials with which the kaon interacts, such as the kapton window of the vacuum chamber and the beam pipe. Among these, the most significant contribution comes from the gas density, which is a key input to the Monte Carlo simulation, as it determines the kaon stopping probability in the neon gas. This uncertainty, linked to the precision of the gas temperature and pressure sensors, is estimated to be $\pm$5\%.

The results shown in Table \ref{tab:KNe} reveal that transitions with $\Delta$n = 1 have a yield reaching up to 30\%. This makes kaonic neon particularly suitable for precision studies of QED, as a high yield is crucial for achieving sub-eV precisions without the need for extremely long data taking.

\section{Conclusions}
Kaonic atoms are unique tools for probing the interactions between kaons and nuclei. The SIDDHARTA-2 collaboration has achieved a significant milestone in measuring the energies of high-n transitions in kaonic neon with high precision, as well as determine their yields, providing input data to develop the theoretical models accounting for the cascade processes in kaonic atoms. This result demonstrate that precision measurements of high-n transitions in kaonic atoms using low-Z gaseous targets are feasible.\\
The first kaonic neon measurement offers new insights into the physics of kaonic atoms, paving the way for a new refined measurement of the kaon mass and of precision tests of BSQED. These results are also relevant to stimulate the interest of theoreticians to perform BSQED calculations for kaonic atoms, in addition to the already established calculations for antiprotonic and muonic atoms. Further investigations and experiments in the field of high precision X-ray kaonic atoms spectroscopy were put forward within the EXKALIBUR (EXtensive Kaonic Atoms research: from LIthium and Beryllium to URanium) project \cite{Curceanu:2023yuy}.

\section*{Acknowledgments}
We thank C. Capoccia from LNF-INFN and H. Schneider, L. Stohwasser, and D. Pristauz-Telsnigg from Stefan Meyer-Institut for their fundamental contribution in designing and building the SIDDHARTA-2 setup. We thank as well the DA$\Phi$NE staff for the excellent working conditions and permanent support. Special thank to Catia Milardi for her continued support and contribution during the data taking. Part of this work was supported by the Austrian Science Fund (FWF): [P24756-N20 and P33037-N] and FWF Doctoral program No. W1252-N27; the Croatian Science Foundation under the project IP-2022-10-3878; the EU STRONG-2020 project (Grant Agreement No. 824093); the EU Horizon 2020 project under the MSCA (Grant Agreement 754496); the Japan Society for the Promotion of Science JSPS KAKENHI Grant No. JP18H05402; the Polish Ministry of Science and Higher Education grant No. 7150/E-338/M/2018 and the Polish National Agency for Academic Exchange( grant no PPN/BIT/2021/1/00037); the EU Horizon 2020 research and innovation programme under project OPSVIO (Grant Agreement No. 101038099). The authors acknowledge support from the SciMat and qLife Priority Research Areas budget under the program Excellence Initiative—Research University at the Jagiellonian University. Catalina Curceanu acknowledge University of Adelaide, where part of this work was done (under the George Southgate fellowship, 2024).

\section*{References}
\bibliography{iopart-num}

\end{document}